\documentstyle[12pt]{article}
\newcommand{\be}{\begin{equation}}
\newcommand{\ee}{\end{equation}}
\newcommand{\bea}{\begin{eqnarray}}
\newcommand{\eea}{\end{eqnarray}}

\newcommand{\V} {{\cal V}_n}
\newcommand{\Y} {{\overline \psi}}
\begin{document}
%%%%%%%%%%%%%Title page%%%%%%%%%%%%%%%%%%%

\begin{center}
\begin{large}
{\bf   Near-Extremal Spherically Symmetric Black Holes \\}
{\bf in an\\}
{\bf  Arbitrary-Dimensional Spacetime \\}
\end{large}  
\end{center}
\vspace*{0.50cm}
\begin{center}
{\sl by\\}
\vspace*{1.00cm}
{\bf A.J.M. Medved\\}
\vspace*{1.00cm}
{\sl
Department of Physics and Theoretical Physics Institute\\
University of Alberta\\
Edmonton, Canada T6G-2J1\\
{[e-mail: amedved@phys.ualberta.ca]}}\\
\end{center}
\bigskip\noindent
\begin{center}
\begin{large}
{\bf
ABSTRACT
}
\end{large}
\end{center}
\vspace*{0.50cm}
\par
\noindent

In a recent paper (hep-th/0111091), the near-extremal thermodynamics 
of a 4-dimensional Reissner-Nordstrom  black hole 
had been considered. In the 
current paper, we extend this  prior treatment to the more general case of
a spherically symmetric, charged black hole of arbitrary  
dimensionality. After summarizing the earlier work, we demonstrate a  
duality that exists between the near-extremal sector of spherically 
symmetric black holes and Jackiw-Teitelboim  theory. On the basis 
of this correspondence, we argue that back-reaction effects prohibit 
any of these ``RN-like''  black holes from reaching extremality and, 
moreover, from coming arbitrarily close to an extremal state.

%PACS 04.70.Dy
\newpage
%\renewcommand{\baselinestretch}{1.5}

%\section{Introduction}
%\medskip
\par

\section{Introduction}

So far,
we have witnessed
 only   
 limited progress  towards  realizing
  a formal  theory of quantum gravity. Yet, there
has  still been much  support for  a fundamental relationship 
between the proposed constituents.    
Nowhere is this relationship more in evidence than
in the thermodynamic behavior of black holes; where gravitation,
the laws of thermodynamics and quantum theory  appear to  been linked together
in some profound, but mysterious manner.
\par
In regard to black hole thermodynamics,  the most prominent
open question is the microscopic origin of black hole entropy.
There have been many attempts at resolving the issue with
varying degrees of success. 
(For a review and references, see Ref.\cite{wald}.)
Perhaps the most profound resolution has been proposed
by Strominger and others \cite{str} in the context
of weakly coupled string theory;  where massive string states 
can be represented
by extremal black holes. In these works, a statistical
procedure has been used to  generate, precisely, the Bekenstein-Hawking area
law (i.e., entropy is given by one quarter of the horizon surface area 
\cite{bek,haw}). 
 It follows, by implication,
that  the extremal limit (i.e., the degeneracy of two otherwise
distinct horizons in a charged or rotating black hole) must
be well defined, with the corresponding
 entropy being  a non-vanishing, mass-dependent quantity.
 Alas, this is in direct conflict with the
Nernst's enthropic formulation of the third law of thermodynamics:
as the temperature of a system approaches zero, the 
  entropy  must approach a constant (typically zero) that is
 independent of all macroscopic parameters of the 
system \cite{thermo}.\footnote{This
formulation of the third law should not be confused
with Nernst's other version: it is impossible
to  reach absolute zero by a finite number of reversible
processes. The two formulations, although typically equivalent,
need not coincide for ``exotic'' thermodynamic systems (such as black holes).}
In contrast to the third law, the  extremal limit corresponds to
 a vanishing   Hawking temperature \cite{haw2} 
 but a Bekenstein-Hawking entropy that (at least naively)
depends on the  mass of the black hole.
\par
In view of this apparent conflict between the third law
 and the  extremal limit of black  hole thermodynamics,
Hawking and others \cite{hhr} have argued as follows.
Extremal black holes
and non-extremal black holes are qualitatively distinct entities,
with no possibility of one class continuously deforming into the other.
In this picture, extremal black holes are indeed assigned zero
entropy, assuming that they can exist at all.
Since the original conjecture, there have been various
other semi-classical calculations  in support of this  viewpoint.
(See Ref.\cite{new} for an up-to-date list of the relevant citations.) 
\par
However, in spite of Hawking et al.'s 
convincing arguments and considerable supporting evidence,
the status of extremal thermodynamics remains, very much, an
open question. This ambiguity can be  attributed to the
 implications
of the forementioned, compelling string-theory calculations.\footnote{For
a list of other papers supporting  a well-defined extremal limit,
see Ref.\cite{me1}.}  Given the apparent incompatibility
of these two  points of view,  any  evidence, one way or the other, 
should be closely examined.  In this spirit,  let us proceed
towards the focus of this paper.
\par
Recently, this author has   considered
the near-extremal thermodynamics of a 4-dimensional Reissner-Nordstrom
(i.e., spherically symmetric and charged) black hole \cite{new}.
Let us briefly review, in point form, the procedure used and the outcomes
of this analysis.
\par
(i) We began by reviewing a duality \cite{fnn} that 
exists between the near-extremal
sector of Reissner-Nordstrom (RN) black holes and 2-dimensional
anti-de Sitter (AdS) gravity.\footnote{It is interesting to note that
this duality  has its origins in the 
AdS$_{2}$/CFT$_{1}$ correspondence 
\cite{x2,x3,x4,x5}.} 
The static black hole solutions
of the latter are described by what is known as
 Jackiw-Teitelboim (JT) theory \cite{jt}.
\par
(ii)
It was shown that
the thermodynamic properties  of a near-extremal RN black hole 
coincide precisely
with near-massless JT thermodynamics.  Significantly, the massless limit
of JT black holes is, at least classically,  
a well-defined procedure for which 
 the associated temperature and entropy both vanish. Hence, from
a classical perspective, this correspondence strongly supports the
viewpoint of a well-defined extremal limit.
\par
(iii)
After reviewing the classical duality, we went on
to examine the corresponding situation at a one-loop level.
In particular, we  incorporated   a (massless) quantum scalar
field into the formalism
and then  considered  the  first-order back reaction on  the JT geometry.
Our focus was on the quantum-corrected surface gravity
(i.e., temperature \cite{haw2}). Inspired by a prior study \cite{aht},
we adopted the viewpoint that a consistent black hole solution
demands a non-negative surface gravity. 
\par
(iv)
Initially, we considered  a  scenario where the quantum matter field
is minimally coupled in the effective 2-dimensional theory.
From this perspective, we found that a massless JT black hole
 remains a well-defined limiting case
(i.e., the surface gravity remains non-negative).
We argued, however, that this scenario is  inappropriate  for the 
following reason. The JT action, in this context, has its origins
in a higher-dimensional theory, and so any matter field should
be subjected to the same criteria.
\par
(v)
With the above argument in mind, we subsequently
considered a matter field with a 4-dimensional pedigree.  That is,
one that is minimally coupled in the originating RN theory.
After repeating the same procedures of dimensional reduction
and field reparametrization that had been imposed on the classical
action, we found that the effective one-loop action mimics
a  dimensionally reduced (non-rotating) BTZ model \cite{btz,ao}.
By directly applying prior works of relevance 
to this model \cite{me1,me2},\footnote{For
earlier studies   on   dimensionally 
reduced theories of this nature,
see \cite{ksol}-\cite{ll}.} we were able to demonstrate that
the  limiting procedure will break down before a vanishing   mass can be
 achieved.
That is to say, the one-loop surface gravity
will only remain non-negative for finite values of the JT black hole
(renormalized) mass. This finite lower bound can be quantitatively expressed
in terms  of  fundamental constants and
the observables (mass and charge) of the dual  RN black hole.\footnote{It
should be pointed out that  one-loop  calculations have  previously
been  carried out for
 a near-extremal, 4-dimensional black hole \cite{msol}.
It was not, however,  the intent of Ref.\cite{new} to
perform rigorous one-loop calculations, but rather to
consider qualitative features in the context of the observed duality.}
\par
(vi)
On the basis of the above result and the observed RN-JT duality, we
concluded that a  RN black hole will be unable to 
achieve extremality. Moreover, the quantum back reaction
will inhibit the RN black hole from even coming arbitrarily
close to an extremal state. Rather, it will ``freeze''
at a finite temperature that is related to the Planck scale.
\par
Let us also briefly touch on another recent study of interest \cite{das}.
Barvinsky, Das and Kunstatter  considered the physical spectra
of charged black holes and deduced that extremal black
holes  can not be achieved (at the quantum level)
due to vacuum fluctuations in the horizon. These authors
also argued that their outcomes applied, quite generically,
to any black hole model that  can  effectively be  described
by a 2-dimensional dilaton theory. 
\par
Given the implied generality of Ref.\cite{das},
the  interest of the current  paper is to ascertain if the results
of our prior analysis \cite{new}  have a more general validity.
In particular, we will  consider a  charged, spherically symmetric black hole 
in an arbitrary-dimensional spacetime. As it turns out, 
the results of our prior study do indeed carry over to
this generalized model.
The rest of this   paper argues in support of this claim.
\par
The proceeding sections are organized as follows. Section
2 introduces the   Einstein-Maxwell action of interest and
considers the near-extremal thermodynamic properties of
the black hole solutions.
In Section 3, we  apply procedures of dimensional
reduction and field reparametrization in obtaining
 an effective Jackiw-Teitelboim theory.
We are then able to demonstrate the duality that exists
between  JT thermodynamics and the near-extremal sector
of the higher-dimensional theory. In Section 4, we
present a simple argument as to why
 the results of the prior study \cite{new} 
persist for this generalized model. Finally, Section 5
contains a brief discussion.

\section{Generalized Einstein-Maxwell Theory}

\par
We  begin with an  $n$+2-dimensional Einstein-Maxwell action:
\be
I^{(n+2)}={1\over 16\pi l^n}\int d^{n+2}x\sqrt{-g^{(n+2)}}\left[
R^{(n+2)}-F^{AB}F_{AB}\right],
\label{1}
\ee
where $l^n$ is the $n$+2-dimensional Newton constant ($l$ being
a length parameter) and
$F_{AB}$ is the Abelian field-strength tensor ($A,B=0,1,...,n+1$). 
\par
The unique static and spherically symmetric solution 
of this action can be described  as follows \cite{huh}:
\be
ds^2=- h(r)dt^2
+{1\over  h(r)}dr^2 +r^2d\Omega^2,
\label{2}
\ee
\be
h(r)=1-{16\pi l^n M\over n \V r^{n-1}}+{32\pi^2 l^{2n} Q^2\over
n(n-1)\V^2 r^{2n-2}},
\label{2.5}
\ee
where $M$ and $Q$ represent the conserved quantities of black hole mass
and  charge (respectively), and $d\Omega^2$ is
an $n$-dimensional constant-curvature hypersurface with
volume  $\V=2\pi^{{n+1\over 2}}/
\Gamma\left({n+1\over 2}\right)$.
\par
If $M^2>nQ^2/2(n-1)$, this is the solution for a charged, 
non-extremal black hole. In this case, the outermost pair
of horizons are distinct and described as follows:
\be
r^{n-1}_{\pm}=
{8\pi l^n\over n\V}\left[M \pm \sqrt{M^2-{n Q^2\over 2(n-1)}}\right].
\label{3}
\ee
\par
\par
Using the usual prescription for non-extremal
black hole thermodynamics \cite{bek,haw,haw2}, we find that
 the associated entropy and temperature  are respectively given by:
\be
S_{BH}={A_{+}\over 4\hbar l^n}= {\V r_+^n\over 4 l^n \hbar},
\label{4}
\ee
\be
T_{H}={\hbar\kappa_+\over 2\pi}= {\hbar\over 4\pi}
\left.{dh\over dr}\right|_{r_+}
= \hbar\left[ {4(n-1)l^nM\over n\V r_+^n}
-{16\pi l^{2n}Q^2\over n\V^2r_+^{2n-1}}\right],
\label{5}
\ee
where $A_+$ is the surface area and $\kappa_+$ is
the surface gravity with respect to the outermost horizon ($r_+$).
\par
For the case of extremal black holes (i.e., 
$r_-=r_+$),  the
associated thermodynamics  remains an open issue (as discussed
earlier). However, we can still safely consider a ``near-extremal
regime'' by setting $\Delta M =M- M_{o}$, where
$M_{o}^2=nQ^2/2(n-1)$ (with the charge 
assumed to be a fixed quantity). Then to leading order in $\sqrt{\Delta M}$: 
\be
r^{n-1}_{+}= {8\pi l^n\over n\V}\left[M_o+\sqrt{2M_o \Delta M}\right].
\label{6}
\ee
Using this expression and  expanding as necessary,
we can obtain near-horizon forms for 
the entropy (\ref{4}) and temperature (\ref{5}).
To leading order in $\sqrt{\Delta M}$, these 
quantities can be expressed as follows (with $w\equiv n-1$):
\bea
\Delta S_{BH}&\equiv& S_{BH}(M,Q)-S_{BH}(M_o)
\nonumber \\
&=& {1\over\hbar} \left[2\right]^{{5\over 4}{n+1\over w}}
\left[\pi\right]^{{n\over w}}  \left[n\right]^{{1\over 4}{n-3\over w}}
\left[w\right]^{-{1\over 4}{5n-3\over w}} \left[\V\right]^{-{1\over w}} 
\left[l\right]^{{n\over w}} 
\left[Q\right]^{{1\over 2}{n+1\over w}}
\sqrt{\Delta M}, 
\label{7}
\eea
\bea
\Delta T_H&\equiv& T_{H}(M,Q)-T_{H}(M_o) 
\nonumber \\
&=& 
\hbar \left[2\right]^{-{1\over 4}{n+9\over w}}
\left[\pi\right]^{-{n\over w}}  \left[n\right]^{-{1\over 4}{n-3\over w}}
\left[w\right]^{{1\over 4}{5n-3\over w}} 
\left[\V\right]^{{1\over w}} \left[l\right]^{-{n\over w}} 
\left[Q\right]^{-{1\over 2} {n+1\over w}}
\sqrt{\Delta M},
\label{8}
\eea
where $S_BH(M_o)\sim M_o^{n\over n-1}$ and $T_H(M_o)=0$.

\section{Effective Jackiw-Teitelboim Theory}
\par
Let us now consider the dimensional reduction of the action (\ref{1})
to an effective 2-dimensional theory.\footnote{For further discussion
on the  various aspects of 2-dimensional gravity, see
 Ref.\cite{strb} and references therein.}
If one imposes the following
ansatz:
\be
ds^2_{n+2}= ds^2(t,x)+ \phi^2(x,t) d\Omega^2,
\label{8.5}
\ee
\be
F_{AB}=F_{\mu\nu}(t,x) \quad where \quad \mu,\nu=0,1\quad only, 
\label{8.75}
\ee
then the following form is obtained (also see Refs.\cite{lk1,newk,x5}):
\be
I= {\V\over 16 \pi l^n}
\int d^2x \sqrt{-g}\phi^{n}
\left[R+ n(n-1)\left({(\nabla \phi)^2\over \phi^2}
+{1\over \phi^2}\right)- {32\pi^2l^{2n}Q^2\over \phi^{2n}\V^2}\right].
\label{9}
\ee
Here,  the ``dilaton'' $\phi$ is identifiable with the radius of the 
symmetric two-sphere,
 the conserved charge is still given by $Q$, and all 
geometric quantities have been  defined with respect to the resultant
 1+1-dimensional
manifold.
\par
At this stage, it is  convenient to redefine
the dilaton as follows:
\be
{\psi}(x,t)= \left[{\phi\over l}\right]^{n\over 2}.
\label{9.5}
\ee
The reduced action (\ref{9}) now reads:
\be
I= {1\over 2G}
\int d^2x \sqrt{-g}
\left[D(\psi) R+ {1\over 2}\left(\nabla \psi\right)^2 
+{1\over l^2} V_Q(\psi) \right],
\label{9.75}
\ee
where we have defined:
\be
{1\over 2G}\equiv {8(n-1)\V\over 16 \pi n},
\label{9.80}
\ee
\be
D(\psi)\equiv {n\over 8(n-1)}\psi^2,
\label{9.90}
\ee
\be
 V_Q(\psi)\equiv
{n^2\over 8}\psi^{{2n-4\over n}}- {4 \pi^2 l^2 n Q^2\over 
(n-1)\V^2 \psi^2}.
\label{9.95}
\ee
\par
The above form of the action is now suitable for
the direct implementation of a field reparametrization 
that is known to eliminate the kinetic term \cite{lk2}.
Thus, we appropriately redefine the dilaton, metric
and ``dilaton potential'' in the following manner:
\be
\Y=D(\psi)={n\over 8(n-1)}\psi^2,
\label{9A}
\ee
\be
{\overline g}_{\mu\nu}= \Omega^2(\psi) g_{\mu\nu}
\label{9B}
\ee
\be
{\overline V}_{Q}(\Y)= {V_{Q}(\psi)\over \Omega^2(\psi)},
\ee
\be
\Omega^2(\psi)= \exp\left[{1\over 2}\int 
{d\psi\over \left(dD/d\psi\right)}\right] = {\cal C}\left[
{8(n-1)\Y\over n}\right]^{{n-1\over n}}.
\label{10}
\ee
Take note of ${\cal C}$, which is a (seemingly) arbitrary constant of 
integration. It can be fixed via physical arguments, and
so we follow Section V of  Ref.\cite{newk} (also see Ref.\cite{x5}) 
and  set: ${\cal C}= n^2/8(n-1)$.
\par
With these reparametrizations, the reduced action (\ref{9.75})  
 takes on the following compact form:
\be
I=\int d^2x \sqrt{-{\overline g}}\left[{\overline \psi} R({\overline g})+ 
 {1\over l^2} {\overline V}_Q({\overline \psi})\right ],
\label{11}
\ee
where we have also set $G=1/2$.
\par
It  is pertinent to this analysis  that the extremal
configuration (i.e., the  extremal limit, 
assuming its existence, in the higher-dimensional model) 
can be recovered
when  ${\overline V}_{Q}({\overline\psi})=0$.
  With the above formalism,  one finds that
 this ``potential'' vanishes  for  ${\overline \psi}={\overline \psi}_o$,
such that:
\be
\left[{\overline \psi}_o\right]^{2{n-1\over n}}
\equiv \left[{2\pi l Q\over (n-1)\V}\right]^2
\left[{n\over 8(n-1)}\right]^{n-2\over n}.  
\label{13}
\ee
\par
With this in mind,
let us now define ${\tilde \psi}\equiv {\overline \psi}-{\overline \psi}_o$
and expand the action (\ref{11}) about the extremal
configuration. To first order in ${\tilde \phi}$, the following is obtained:
\be
I=\int d^2x \sqrt{-{\overline g}}\left[{\tilde \psi} R({\overline g})+ 
 {1\over l^2} {\tilde V}_Q({\tilde \psi})\right ],
\label{14}
\ee
where:
\bea
{\tilde V}_Q({\tilde\psi})&\equiv&\left.
{d {\overline V}_Q \over d{\overline\psi}}\right|_{{\overline\psi}_o}
{\tilde \psi} = 2{(n-1)^2\over n}\left[{n\over 8(n-1)}\right]^{1\over n}
\left[{\overline\psi}_o\right]^{-{n+1\over n}} {\tilde \psi}
\nonumber \\
&=& 2 {(n-1)^2\over n}\left[{8(n-1)\over n}\right]^{{1\over 2}{n-3\over n-1}}
\left[{(n-1)\V\over 2 \pi l |Q|}\right]^{{n+1\over n-1}} {\tilde \psi}. 
\label{15}
\eea
\par
Henceforth, we drop the tildes and bars;  thus considering
the following action:
\be
I=\int d^2x \sqrt{- g}\psi\left[ R( g)+ 
 2{\lambda\over l^2} \right ],
\label{16}
\ee
where $\psi\lambda$ corresponds to one half of  the 
right-hand side of Eq.(\ref{15}).
This is simply the action for 2-dimensional  AdS gravity; the
 black hole solutions of which are the well-known JT black holes \cite{jt}.
\par
It can be readily shown that, for a static gauge,
the general  solution of the JT action (\ref{16})
can be expressed as follows:
\be
ds^2= -(\lambda {x^2\over l^2}-lm)dt^2+(\lambda {x^2\over l^2}-lm)^{-1}
dx^2,
\label{17}
\ee
\be
\psi={x\over l},
\label{18}
\ee
where $m$ represents the conserved mass of the JT black hole.
Moreover, with straightforward application  of Ref.\cite{lk3}
(applicable to  a generic 2-dimensional dilaton  theory),
we are able to identify the following thermodynamic properties:
\be
S_{JT}={4\pi\over \hbar} \psi_{+},
\label{19}
\ee
\be
T_{JT}={\hbar\lambda\over 2\pi l}\psi_{+},
\label{20}
\ee
where $\psi_{+} = x_{+} /  l =\sqrt{lm / \lambda}$
is the horizon value of the dilaton field.
\par
Substituting for $\lambda$ into the above expressions
and also identifying  $m$ with $\Delta M$, we 
 ultimately find the following (with $w=n-1$):
\be
S_{JT}= 
{1\over\hbar} \left[2\right]^{{1\over 4}{7n+3\over w}}
\left[\pi\right]^{{1\over 4}{6n-2\over w}}  
\left[n\right]^{{1\over 4}{3n-5\over w}}
\left[w\right]^{-{1\over 4}{7n-5\over w}} 
\left[\V\right]^{-{1\over 2}{n+1\over w}} 
\left[l\right]^{n\over w} 
\left[Q\right]^{{1\over 2} {n+1\over w}}
\sqrt{\Delta M}, 
\label{21}
\ee
\be
T_{JT}= \hbar \left[2\right]^{-{1\over 4}{3n+7\over w}}
\left[\pi\right]^{-{1\over 4}{6n-2\over w}}  
\left[n\right]^{-{1\over 4}{3n-5\over w}}
\left[ w\right]^{{1\over 4}{7n-5\over w}} 
\left[\V\right]^{{1\over 2}{n+1\over w}} 
\left[l\right]^{-{n\over w}} 
\left[Q\right]^{-{1\over 2} {n+1\over w}}
\sqrt{\Delta M}. 
\label{22}
\ee
A direct comparison of these expressions with Eqs.(\ref{7},\ref{8}) yields
the following intriguing  outcomes:
\be
S_{JT}={\cal K}\Delta S_{BH},
\label{23}
\ee
\be
T_{JT}={1\over {\cal K}}\Delta T_{H},
\label{24}
\ee
where ${\cal K}$ is some dimensionless numerical factor.
\par
It is easy to verify that ${\cal K}=1$
for $n=2$ (i.e., the 4-dimensional RN black hole). 
In general, however, one finds that  ${\cal K}\neq 1$.\footnote{That
4-dimensional gravity holds a privileged position 
 is not all together new. For
example, black holes only saturate a Bekenstein-like entropy
bound in four dimensions of spacetime \cite{bousso}.}
  For instance,
if $n=3$, then ${\cal K}=\sqrt{3/2\pi}$.  However, this
lack of an exact coincidence (when $n\neq2$) is essentially
irrelevant, given that we are always free to  rescale
the fundamental constants  from the perspective of
the lower-dimensional theory. That is,  as far as the
second law of thermodynamics is concerned,  the above thermodynamics
does indeed coincide. (The pertinent points being that $S_{JT}T_{JT}=
\Delta S_{BH}\Delta T_{H}$ and all dimensional quantities
coincide exactly.) Hence, we have demonstrated  the anticipated duality
for any choice of $n$:
the  near-extremal sector of charged,  spherically symmetric black 
holes  with the  near-massless sector of JT theory.

\section{One-Loop Considerations} 
\par
In analogy to the $n=2$ analysis of Ref.\cite{new},
it is also necessary to consider the  implications (to first-perturbative
order) of
a   quantum scalar field. First note that any consideration of
a minimally coupled matter field in two dimensions follows trivially from the
prior treatment.\footnote{This is because  the lower-dimensional,
reparametrized action (\ref{16})  is
formally identical for all $n$.}   However, it is still necessary to examine
the repercussions of a matter field that has its origins in the
higher-dimensional theory.
\par
Let us thus consider a massless scalar field ($f$) that
is minimally coupled with respect to  the original $n$+2-dimensional
Einstein-Maxwell model.  The revised
(total) action   can now be written as:
\be
I^{(n+2)}_{TOT}=I^{(n+2)}-{\hbar\over 16\pi l^n}\int d^{n+2}x\sqrt{-g^{(n+2)}}
(\nabla^{(n+2)}f)^2,
\label{32}
\ee
where $I^{(n+2)}$ is the classically defined  action of Eq.(\ref{1}). 
Again imposing the  spherically symmetric ansatz of 
Eqs.(\ref{8.5},\ref{8.75}) (along
with $f=f(t,x)$), we obtain the following   reduced 
formulation:
\be
I_{TOT}=I-{\hbar\V\over 16\pi l^n}\int d^2 x\sqrt{-g}
\phi^n (\nabla f)^2,
\label{33}
\ee
where $I$ is the reduced action of Eq.(\ref{9}).
\par
Following the exact same pattern of field reparametrization and expansion as
previously described, we ultimately find that:
\be
I_{TOT}=I_{JT}-{\hbar(n-1)\V\over 2\pi n}\int d^2 x\sqrt{-{\overline g}}
{\tilde \psi}
({\overline\nabla}f)^2,
\ee
where $I_{JT}$  is the JT action of Eq.(\ref{14}) (or Eq.(\ref{16})),
and we have explicitly shown the tilde and bar notation  
for the sake of clarity.
\par
As was also  found for the special case of $n=2$ \cite{new}, 
the dilaton-matter coupling  is precisely
that obtained in the dimensional reduction (from three to two dimensions) 
of a BTZ black hole;  assuming minimal coupling in the higher-dimensional
theory \cite{btz,ao}.\footnote{The unorthodox constant factor
in front of the above matter action is irrelevant, as this can always be
absorbed through a redefinition of $f$.}
This means that all of the  outcomes of the prior ($n=2$) analysis 
will automatically  persist for the generic-$n$ case  and need not be 
repeated. Hence, we  conclude that, by way of duality,
an $n$+2-dimensional spherically symmetric, charged black hole
can not reach a state of extremality and, moreover, can not
even come arbitrarily close to an extremal state.

\section{Conclusion} 
\par
In summary, we have extended the results of a prior analysis \cite{new}
on the 4-dimensional Reissner-Nordstrom black hole to
analogous black holes of arbitrary dimensionality. The outcomes,
now applicable to any dimension, argue against the existence
of a well-defined extremal limit.  The basis of these
arguments is that a consistent black hole solution 
requires a non-negative surface gravity \cite{aht}.  With this criteria,  it 
can readily be verified
\cite{new} that an effective (2-dimensional) Jackiw-Teitelboim theory 
\cite{jt} 
must have a lower bound imposed on  its black hole mass.   
By way of  a duality (which we have clearly demonstrated),
it follows that  $n$+2-dimensional RN-like black holes
will ``freeze'' at some finite temperature before
ever reaching extremality.
\par
 It is interesting to note
that the same treatment can be readily applied to a  rotating BTZ
model \cite{btz}, with equivalent outcomes. 
This follows, almost trivially, by  virtue of the
known duality (between the near-extremal BTZ sector
and JT theory \cite{x4}) and  simplifications that
are inherent to the BTZ analysis (primarily, the absence
of a kinetic term in the higher-dimensional action).
It is not yet clear, however, how more ``exotic'' 
black hole geometries may hold up; for instance,
$n$+2-dimensional Reissner-Nordstrom-anti-de Sitter
black holes.  On the other hand, given the wide class of black holes
exhibiting a near-extremal duality with AdS$_2$ \cite{wee},
we anticipate that  similar outcomes are obtainable for
many other black hole scenarios. We will  defer this
question to future studies.  

\section{Acknowledgments}
\par
The author would like to thank V.P. Frolov for helpful conversations.
\par

\end{document}